# Tonal Frequencies, Consonance, Dissonance: A Math-Bio Intersection

Steve Mathew[1,2*]

1. Indian Institute of Technology Madras, Chennai, IN

2. Vellore Institute of Technology, Vellore, IN

*Corresponding Author

Email: stevemathewda@gmail.com

**Abstract**

To date, calculating the frequencies of musical notes requires one to know the frequency of some reference note. In this study, first-order ordinary differential equations are used to arrive at a mathematical model to determine tonal frequencies using their respective note indices. In the next part of the study, an analysis that is based on the fundamental musical frequencies is conducted to theoretically and neurobiologically explain the consonance and dissonance caused by the different musical notes in the chromatic scale which is based on the fact that systematic patterns of sound invoke pleasure. The reason behind the richness of harmony and the sonic interference and degree of consonance in musical chords are discussed. Since a human mind analyses everything relatively, anything other than the most consonant notes sounds dissonant. In conclusion, the study explains clearly why musical notes and in toto, *music* sounds the way it does.

**Keywords**: Cognitive Musicology, Consonance, Mathematical Modelling, Music Analysis, Tension.

## 1. Introduction

The frequencies of musical notes follow an exponential increase on the keyboard. It is the basis upon which we choose to use the first-order ordinary differential equations to draw the model. In the next part of the study, we analyze the frequencies using sine waves by plotting them against each other and look for patterns that fundamentally unravel the reason why a few notes of a chromatic scale harmonize with the tonic and some disharmonize. It was expected that a concrete conclusion could be drawn from the graphs. Au contraire, we could only group them into similar notes. Then we look at a way to further refine the problem by considering how our brain perceives music. The parietal lobe is responsible for the perception of music *(1)*, yet while listening to music many areas of the brain are activated *(2)*. Using the fundamental principles of neurobiology and music, we will then understand the formation of scales and sweet-sounding patterns of sound. In continuation, we look at the reason behind the working of some fundamental principles of harmony and conclude the reason for the richness of harmony. With the same understanding of musical notes, we then look at the working of chords and analyze the property of dissonance and the degree of consonance in them to conclude why we hear what we hear, the way it is heard.

Previous studies proving musical properties include:

In *Mathematics and Music*, David Wright states that the frequencies of all the other notes of a piano can be calculated by using the note A4 (440 Hz) as a reference and the frequencies of consecutive semitones differ by a factor of the 12th root of 2 *(3)*. The human brain is highly capable of recognizing systematic auditory patterns *(4-6)*. The reason behind perceived consonance and dissonance were also analysed by relating them with human speech sounds *(7-10)*. The unison, fourth, fifth and octave were proved to be consonant which goes in agreement with the results of this study as well *(11)*. Leibnitz recognized that the numerical ratios of frequencies are not consciously perceived rather they were the results of unconscious calculations *(12)*. This is a fundamental principle that acts as a bridge linking the mathematical and neurobiological considerations of this study and supports the findings. Smoothness of a tone when heard has led to distinguishing consonance and dissonance; Helmholtz used smoothness of tones as a parameter for making decisions of consonance saying, "consonance is a continuous, dissonance an intermittent sensation of tone. Two consonant tones flow on smoothly . . . dissonant tones cut one another up into separate pulses of tone" *(13)*. Analysing consonance and dissonance using dyads have been quite popular *(14)*. In this study, the same is used to plot the sinusoidal waves for frequencies for a duration of time. Studies have been engaged in ranking the musical intervals based on the degree of consonance and to evaluate the ability to perceive consonance in terms of this ranking *(13)*. The concept of virtual pitch has also been a base for studies on consonance *(15)*.

## 2. Methods

### *2.1 A Mathematical Model to determine frequencies of musical notes*

The following study is conducted concerning a grand piano (A440) to relate to the sounds that are well within the human hearing range *(16,17)*. However, the results are also applicable to the infinite notes that are present outside the human hearing range. This study was based on an assumption that the change in frequencies of musical notes with respect to the number of notes is directly proportional to the total frequency. Here we use the already known values *(18)* of frequencies of musical notes.

Here 'x' denotes the frequency of a musical note and 'y' denotes the note order (from 1 to 88)

$\frac{dx}{dy} \propto x$

By separating the variables and integrating on both sides,

$x = ce^{ky}$ , here c is an arbitrary constant and k is the constant of proportionality.

By taking the values of A2 (x=110 Hz and y=25) and A#4 (x=466.16 Hz and y=50), we arrive at the expression,

$$x = 25.956753046164 * e^{0.0577619426421*y}$$

The above expression follows simplification using basic algebra and logarithm.

*2.2 Consonance and Dissonance: Scales*

The following study is conducted based on the hypothesis that the degree of consonance of a musical note with respect to its key-center is determined by the periodicity and closeness of intersections of the waves plotted using the formula y(t) = sin (2πft) *(19)* (where f is the tonal frequency and t is the time from the origin of sound) of the key-center, the note in analysis and the axis where $\sin^{-1}(t) = 0$. The points in time where they intersected as expected are tabulated (Table. 1). The recordings were taken in the D Chromatic Scale where the conclusions apply to every single scale. It is to be considered that the values that are recorded were accurate to 3 decimal places and absolutely perfect intersections could not be recorded owing to the minute variations as the tonal frequencies are primarily based on disturbance of air molecules. Another complexity in this study is that sometimes the analysis gets interchanged into multiple scales i.e. The graphs of the tonic and supertonic may seem promising substantiating consonance; the same way, the tonic and minor $7^{th}$ might exhibit the same. This is because in the major scale of the current minor $7^{th}$, the tonic in discussion is diatonic but the converse is not true. Thus graphically, the minor 2nd is similar to major $7^{th}$, the minor $7^{th}$ is similar to supertonic, the minor $6^{th}$ is similar to mediant, the minor $3^{rd}$ is similar to sub-mediant. Looking at the notes of the D diatonic major scale, the supertonic exhibits intricate patterns. In this case, it shows the intervals ~0.449, ~0.109, ~0.109 repeatedly. A similar pattern is found in the minor $7^{th}$ (though the intervals change). The mediant has intervals shuffled between 2 differences. This shuffle of intervals between 2 set values of differences is found in minor $6^{th}$ as well. The previous applies for the minor $3^{rd}$ and sub-mediant too. These show that there are "mutual" relationships between the graphically similar notes with the current tonic (the note D) being the link. The notes sub-dominant and dominant intersect perfectly graphically i.e., periodic intersections proving to be consonant which goes in agreement with the previous studies *(15)*. The tritone is quite unstable as it does not have a common difference in its intersections. The statement of irregular intersections can be contradictory to the fact that sinusoidal waves are periodic. Yet this is cleared by the fact that the recordings are only accurate to 3 decimal places. The sinusoidal intersections would be periodic if the intersections are recorded by anticipating their occurrence at periodic intervals which would also maintain the uniformity in the error of each recording; the errors would be whole number multiples of the first error. Still, when all the recordings accurate to 3 decimal places were considered, there were intricate, systematic patterns of intersections as tabulated (Table. 1).

**Table 1** Points of intersections of the respective sine graphs of tonic and other notes of the chromatic scale.

| Notes in the Chromatic Scale | Points on the graph where the intersections take place (Time in seconds) | | | | | | | | | |
|---|---|---|---|---|---|---|---|---|---|---|
| **Supertonic** | 0.1091 | 0.5583 | 0.6674 | 0.7766 | 1.2256 | 1.3348 | 1.444 | 1.893 | 2.0022 | 2.11144 |
| **Mediant** | 0.05411 | 0.3135 | 0.36757 | 0.4216 | 0.681 | 0.9406 | 0.9946 | 1.0487 | 1.308 | 1.3621 |
| **Sub -Dominant** | 0.04087 | 0.0816 | 0.12242 | 0.1631 | 0.2042 | 0.2448 | 0.2858 | 0.3264 | 0.3674 | 0.4081 |
| **Dominant** | 0.0273 | 0.0545 | 0.08181 | 0.1091 | 0.1363 | 0.1636 | 0.1908 | 0.2181 | 0.2454 | 0.2727 |
| **Sub-Mediant** | 0.2591 | 0.29966 | 0.3402 | 0.5588 | 0.5993 | 0.6398 | 0.8584 | 0.8989 | 0.9394 | 1.1582 |
| **Leading Tone** | 0.2453 | 0.4906 | 0.7216 | 0.9668 | 1.2121 | 1.457 | 1.702 | 1.948 | 2.179 | 2.4243 |
| **Minor Second** | 0.2314 | 0.4628 | 0.6821 | 0.9128 | 1.1442 | 1.375 | 1.607 | 1.838 | 2.056 | 2.2884 |
| **Minor Third** | 0.2176 | 0.504 | 0.7902 | 1.0079 | 1.2942 | 1.511 | 1.798 | 2.0158 | 2.233 | 2.5198 |

| | | | | | | | | | |
|---|---|---|---|---|---|---|---|---|---|
| Tritone | 0.1637 | 0.2312 | 0.3948 | 0.5586 | 0.7897 | 0.9535 | 1.1846 | 1.348 | 1.512 | 1.743 |
| Minor sixth | 0.06862 | 0.1631 | 0.2317 | 0.3002 | 0.3947 | 0.4633 | 0.5578 | 0.6264 | 0.695 | 0.7894 |
| Minor Seventh | 0.12233 | 0.6268 | 0.7491 | 0.8714 | 1.375 | 1.498 | 1.62 | 2.125 | 2.247 | 2.369 |

When 2 graphs are similar, it can be quite confusing as to which one gets its way into the scale/tagged consonant. Keeping the tonic, sub-dominant and dominant constant and experimenting with the vulnerable notes, we see that we have a variety of options. Systematic sounds make our brains expect what comes next owing to the patterns our brains have already registered *(2)*. Thinking of this neurobiological ground rule, we can try to analyze the structure of the major scale and interpret it accordingly. In the major scale, we know that the intervals between the 3$^{rd}$ and 4$^{th}$ notes and the 7$^{th}$ and 8$^{th}$ notes are semitones and the remaining are whole-tones *(3)*. By drawing an imaginary line between the 4$^{th}$ and 5$^{th}$ notes, we can divide the scale into 2. We can see that the intervals are in the sequence, 'tone, tone, semitone' in both the segments. When the various permutations are analyzed, there are only 4 possibilities where the final note of the scale resolves back into the tonic where the same pattern of intervals are found in both the segments. They are the Ionian, the Phrygian, the Dorian and the Double Harmonic Major. Even if we come up with various other "systematic" scales (with lesser number of notes), they would still be excerpts of the above scales. It is also surprising that when we see the graphs of notes that look similar because of their mutual relationships, the notes can be interchanged to swap between Ionian and Phrygian. It is also intriguing that the 4 altered notes are super-tonic, mediant, leading tone and sub-mediant which are the root notes for the formation of minor and diminished chords diatonic to the scale. Thus, those notes have an immense effect on changing the entire mood of the scale. The Dorian scale can be considered as a scale that partially resolves the darkness created by the Phrygian scale (similar to the way leading tone resolves to tonic) reducing the overall tension of the scale. This is the reason Dorian does not sound as bright to the musically conscious brain as Ionian but brighter than Phrygian *(20)*. This analysis also proves that consonance and dissonance are related to the scale. They both can also be felt when the two notes are played individually and consecutively apart from the subject of disharmony when they are played together. This is because our brains read auditory signals relatively *(21)* and every time we hear a note as a part of a sequence of notes, our brain tries to associate it with the tonic *(22,23)*. This neurological characteristic where subconsciously our brains identify patterns suggests that our brain chooses Ionian, Lydian and Phrygian over the others. The Lydian is still considered to be the brightest scale, yet it does not feel predictable owing to its 'not so' systematic patterns. Another scale that has this systematic sequence of notes is the double harmonic major scale *(24)*. Though this is a scale with a systematic pattern, this does not sound bright, as it has a 1.5 tone (large) interval between its 2$^{nd}$,3$^{rd}$ and 6$^{th}$, 7$^{th}$ degrees. This is partially found in the harmonic minor scale; only between 6$^{th}$ and 7$^{th}$ degrees *(25)*.

*2.3 Stability of a Harmonic Couple*

When harmonizing a piece of music vocally or instrumentally, it is often done by making the performers start in different notes of the same scale. It is common to harmonize using the mediant *(25)* and the dominant as it follows the major triad of the tonic. We know that the 5$^{th}$ is a stable note in the scale and there are 6 common notes in the diatonic major scales of the current tonic and the current dominant. So, this seems a lot less tense and is seen to have a smooth transition in the melody. Speaking of harmonizing using the mediant, we see that the tonic and the mediant have only 3 notes in common in their respective diatonic major scales; which means that for a performer to harmonize by starting with the mediant, the performer will need to go outside the scale of the mediant more than half the time when required to produce the remaining 4 notes that are diatonic only to the tonic of the "main performer". This implies that the musicians here play in a pattern that follows the 'similar motion' *(25) as in contrapuntal music.* The underlying beauty of going "off-course" from the scale (with respect to the harmonizer's tonic) and then resolving gives the edge for the mediant when harmonizing with the tonic which is not the same for the dominant where the need to go off-course is not so much even while exhibiting similar motion. The root of the harmonizer's melody, the mediant itself is a vulnerable note and is not very stable. This lack of attainable stability makes harmony richer. In defense, there is a pattern that is found in the scale formed by the similar motion from the harmonizer's end. When seen closely, we can see that, the scale of the harmonizer (starting from the mediant) corresponding to the main performer is in the Phrygian scale with the mediant as the tonic, at the same time when the main performer plays in the Ionian of the tonic in the discussion. This blend of bright and dark, gloomy scales characterized by their accommodation of specific notes in a systematic arrangement makes the harmony appealing and rich. Thus, it would be pleasing to infer that *the lack of stability of notes makes the harmony richer.*

### *2.4 Consonance and Dissonance: Chords*

Interference of sounds occurs in chords. They can interfere either way to form consonance or dissonance. An accidental is any note that is foreign to the scale we are then currently in. It would be pleasing to propose that *'The effect of an accidental is determined by the other notes that are played along with it in a chord.'* Thinking about the same with an example; it can be observed that in a major scale, the minor chord of the sub-dominant might sound extremely dissonant. It is obvious that it houses the minor $6^{th}$ which is significantly responsible for that effect. But playing the Major chord of the mediant as a passing chord between the tonic and the minor of the sub-mediant (relative minor) does not sound so dissonant. The reason for the same can be found by observing the Table 1 closely; It is seen that at 0.1632 seconds (shown in Figure. 1), the sub-dominant and minor $6^{th}$ interfere with each other when they are played with the tonic. This makes their sound more distinct, yet highly dissonant as the disharmonizing effect of minor $6^{th}$ is given a lift. This case is not found in the major chord of mediant making it a lot less dissonant. Using the same values of Table. 1, the reason for the consonance of triads were found. It would be pleasing to propose that the consonance of triads/chords are characterized by the quickness in which they prove to be in sync with the tonic. This is found by analyzing the values of notes that form the various possible triads. By taking a look at the values of mediant and dominant, we try to find a common intersection (accurate to 2 decimal digits) between them i.e. the point in time where all the 3 notes of the triad are in consonance. The triad that takes the shortest time to have all of its note in consonance is the most consonant. The most consonant seems to be the major triad at 0.054 seconds, the sus 4 chord at 0.081 seconds, the sus 2 chord at 0.1091 seconds, the minor triad at 0.21 seconds, the augmented triad at 1.41 seconds and the diminished triad at 1.51 seconds. This ascending order of times goes in agreement with the degree of consonance perceived. The diminished triad houses the tritone which is unsettling and is self-explanatory on the topics of dissonance in both diminished triad and tritone.

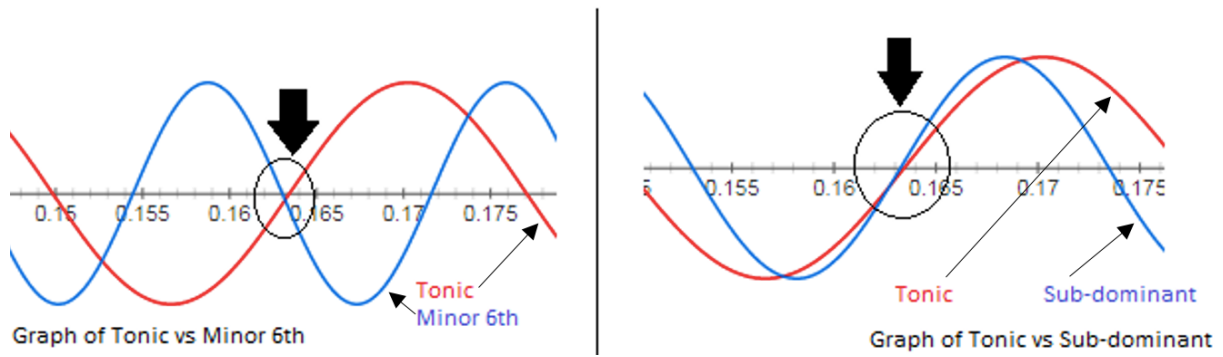

**Figure. 1.** Intersections of minor 6[th] and sub-dominant with the tonic.

## 3. Results

In this study, we arrive at a mathematical model to determine the frequencies of musical notes, followed by an analysis based on fundamental tonal frequencies providing a wider understanding of the scales and notes and the reason behind its harmony or disharmony. The further refining of results was done by looking at the problem from a neurobiological perspective and the same was achieved. The same was applied to understand the musicality when two or more notes were played simultaneously, and the results concluded how the notes when used in different scenarios sound differently.

**Supplementary Text**

Steps to arrive at the mathematical model:

$x = ce^{ky}$ (Equation 1) has two unknowns c and k. To find c and k, few values of known x and y are substituted.

Frequency of A2 = 110,
x = 110, y = 25
=) $110 = ce^{25k}$                                    Equation 2

Frequency of B♭4 = 466.16Hz
x = 466.16, y = 50
=) $466.16 = ce^{50k}$                                Equation 3

From Equation 2, $c = \frac{110}{e^{(25k)}}$

From Equation 3, $c = \frac{466.16}{e^{(50k)}}$

Equating both, $\frac{110}{e^{(25k)}} = \frac{466.16}{e^{(50k)}}$

=) $\frac{110}{466.16} = \frac{1}{e^{(25k)}}$

$e^{(25k)} = 4.23781822$

25k = ln (4.23781822)

25k = 1.4440485660522

k = 0.0577619426421

Now that we've got the value of k, we substitute it as follows,
=) $x = c * e^{0.577619426421*y}$

From Equations 2,3:
$110 = ce^{(25k)}$
$466.16 = ce^{(50k)}$

From Equation 2,
$110^2 = c^2 e^{(50k)}$
$12100 = c (ce^{(50k)})$
12100 = c (466.16)
c = 25.956753046164

Now we substitute c as follows:
=) $x = 25.956753046164 * e^{0.0577619426421*y}$